\documentclass[conference]{IEEEtran}
\IEEEoverridecommandlockouts
\usepackage{cite}
\usepackage{amsmath,amssymb,amsfonts}
\usepackage{algorithmic}
\usepackage{graphicx}
\usepackage{textcomp}
\usepackage{xcolor}
\def\BibTeX{{\rm B\kern-.05em{\sc i\kern-.025em b}\kern-.08em
    T\kern-.1667em\lower.7ex\hbox{E}\kern-.125emX}}
\begin{document}

\title{Comprehensive Modeling of Camera Spectral and Color Behavior\\
\thanks{Royal Society of New Zealand - Rutherford Discovery Fellowship}
}

\author{\IEEEauthorblockN{Sanush K Abeysekera}
\IEEEauthorblockA{\textit{School of Engineering} \\
\textit{University of Waikato}\\
Hamilton, New Zealand \\
0000-0002-0737-9418}
\and
\IEEEauthorblockN{Ye Chow Kuang}
\IEEEauthorblockA{\textit{School of Engineering} \\
\textit{University of Waikato}\\
Hamilton, New Zealand \\
0000-0002-5423-9653}
\and
\IEEEauthorblockN{Melanie Po-Leen Ooi}
\IEEEauthorblockA{\textit{School of Engineering} \\
\textit{University of Waikato}\\
Hamilton, New Zealand \\
0000-0002-1623-0105}
}

\maketitle

\begin{abstract}
The spectral response of a digital camera defines the mapping between scene radiance and pixel intensity. Despite its critical importance, there is currently no comprehensive model that considers the end-to-end interaction between light input and pixel intensity output. This paper introduces a novel technique to model the spectral response of an RGB digital camera, addressing this gap. Such models are indispensable for applications requiring accurate color and spectral data interpretation. The proposed model is tested across diverse imaging scenarios by varying illumination conditions and is validated against experimental data. Results demonstrate its effectiveness in improving color fidelity and spectral accuracy, with significant implications for applications in machine vision, remote sensing, and spectral imaging. This approach offers a powerful tool for optimizing camera systems in scientific, industrial, and creative domains where spectral precision is paramount.
\end{abstract}

\begin{IEEEkeywords}
cameras, sensitivity, modeling
\end{IEEEkeywords}

\section{Introduction}
Digital cameras have become indispensable tools in various fields, ranging from photography and industrial machine vision to scientific imaging and remote sensing. At the heart of these devices lies the spectral response of the camera, which defines how scene radiance is transformed into pixel intensities for each color channel. Understanding and accurately modeling this response is critical for applications where color fidelity and spectral accuracy are paramount.

\begin{figure}[h]
\includegraphics[width=\columnwidth]{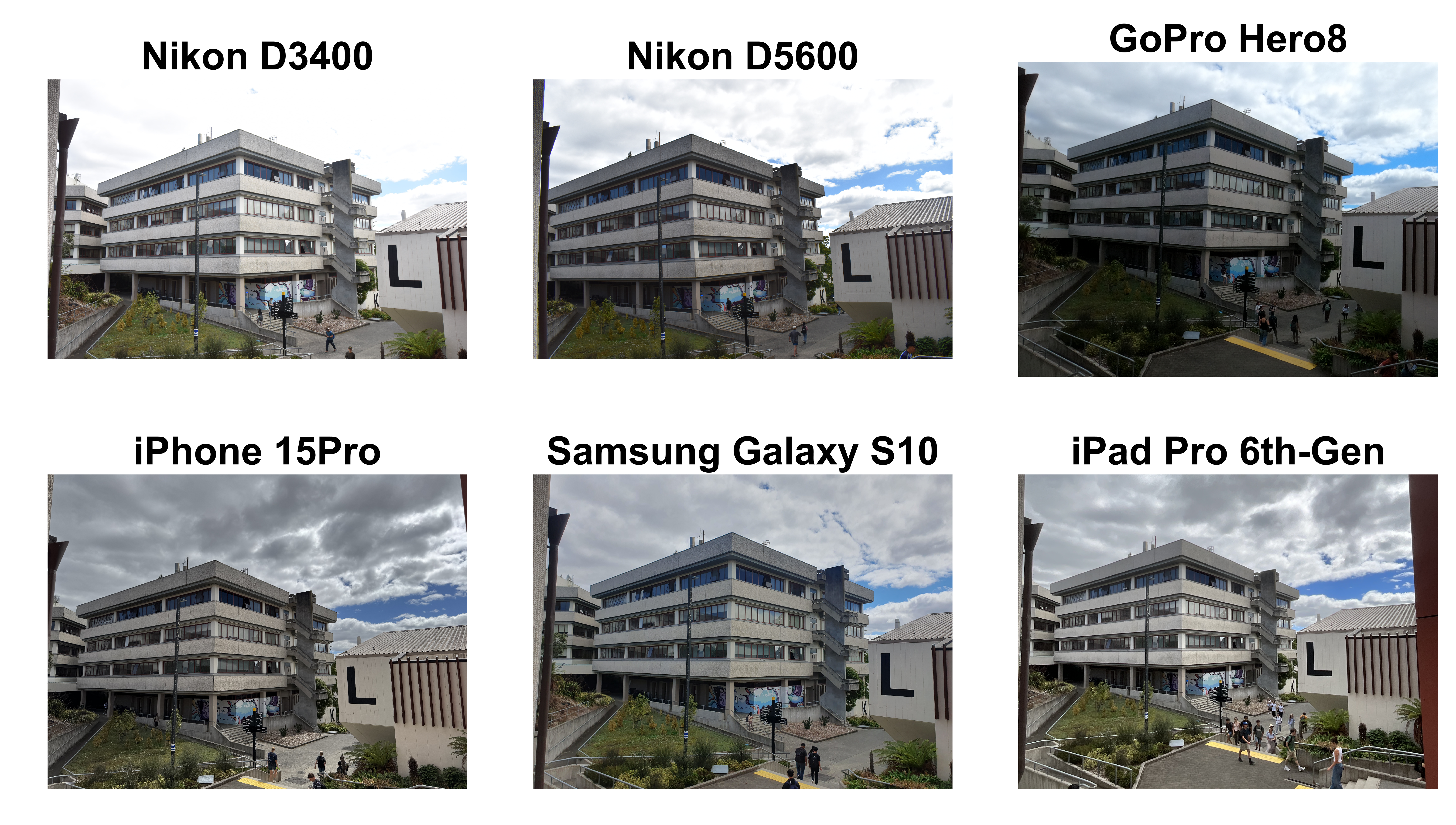}
\caption{Images of the same scene captured by 6 different cameras.}
\label{fig1}
\end{figure}

To our knowledge, no single existing work has constructed a comprehensive, end-to-end model that links incoming illumination to output pixel intensity. In our previous work, we have observed the importance of such a model for ensuring cross-system consistency among cameras. For example, as illustrated in \ref{fig1}, the same scene captured by six different cameras exhibits varying levels of brightness. This variation suggests that a computer vision algorithm optimized for one camera may require retraining to maintain performance when paired with a different camera. However, the end-to-end camera model presented in this paper allows computer vision algorithms to interface seamlessly with new cameras, minimizing the need for retraining and ensuring consistent performance across diverse camera systems.

In practice, a camera’s ability to detect scene radiance at the lens is influenced by several complex factors, including optical distortions such as vignetting and lens fall-off \cite{rodrigues2020, grossberg2004, asada1996}.  Additionally, image acquisition introduces various sources of noise, including dark noise, fixed pattern noise, and shot noise \cite{grossberg2004}, as well as interference from adjacent pixels \cite{klein1996}. However, since this paper focuses on modeling the internal functioning of a camera, these noise sources are not considered in the proposed model.

\begin{figure}[h]
\includegraphics[width=\columnwidth]{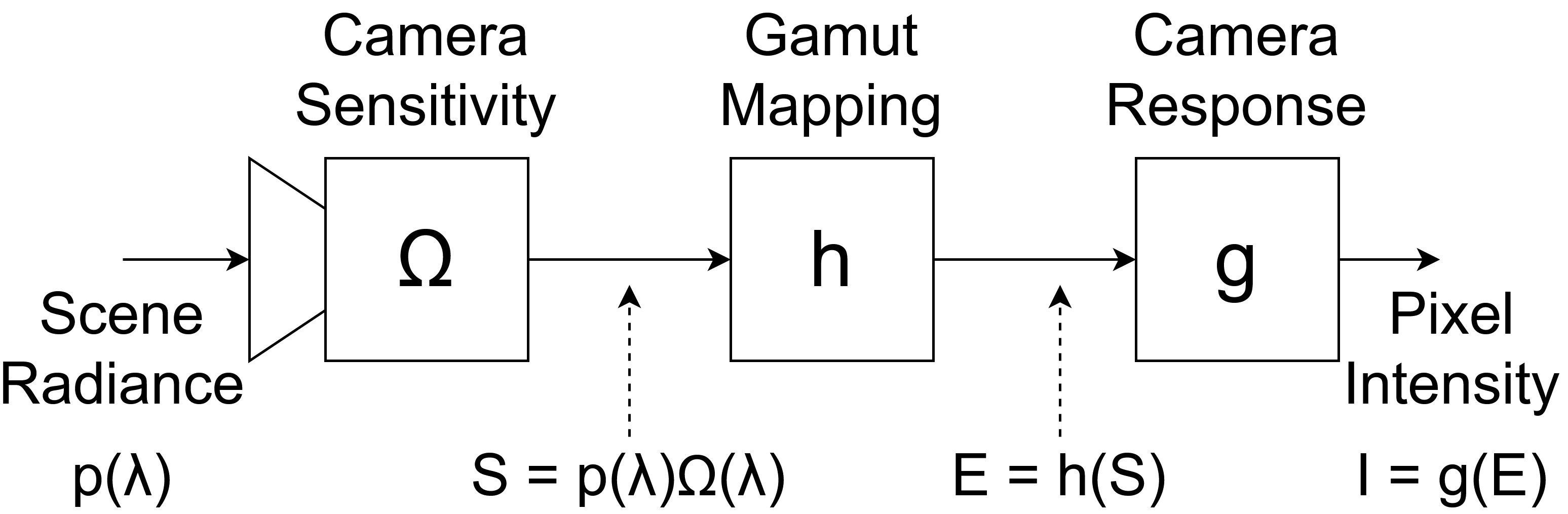}
\caption{The proposed image acquisition pipeline shows the interface between scene radiance and pixel intensity. Here, $P(\lambda)$ indicates that $P$ is a function of wavelength $\lambda$.}
\label{fig2}
\end{figure}

The image acquisition pipeline used in this study is depicted in \ref{fig2}. In the interest of both accuracy and convenience, we present only the most critical steps of the pipeline. Alternative pipelines at varying levels of completeness are provided in \cite{debevec1997,kim2012}.

\begin{figure*}[t]
\includegraphics[width=\textwidth]{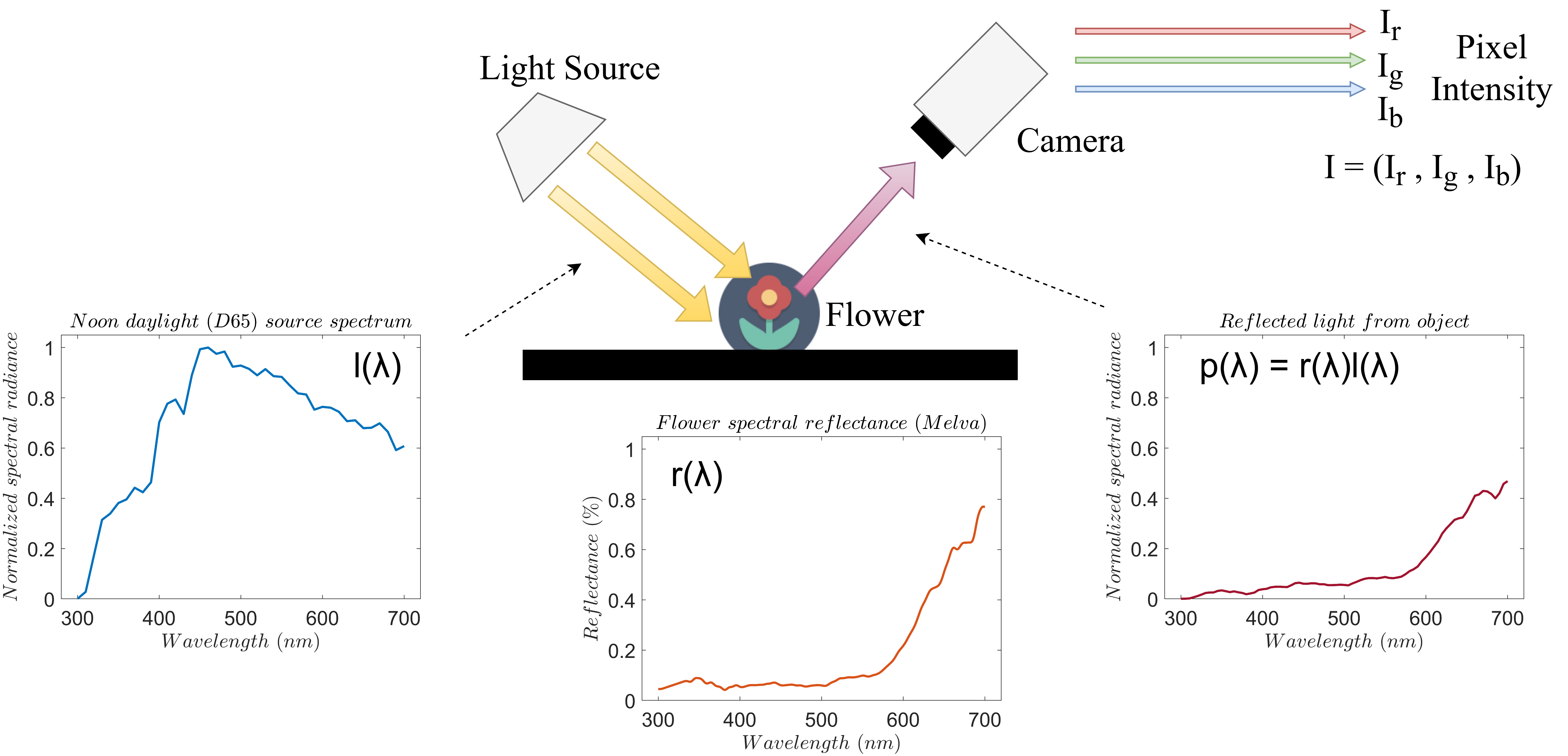}
\caption{Physics of Image Acquisition. The scene radiance perceived by the camera results from the combination of illumination incident on the object and the object’s reflectance. The output of an RGB camera is a red, green, and blue pixel intensity triplet.}
\label{fig3}
\end{figure*}

As noted above, no single investigation has presented an end-to-end model that links scene radiance and pixel intensity. However, individual components of the camera pipeline have been studied in great detail over the years. In fact, this paper serves as an interface between these existing models, integrating them to create a single holistic framework. To enhance the ease of implementation, particular attention was given to developing a modeling procedure that does not require specialized optical equipment. A key piece of equipment in the proposed data acquisition stage is a custom-designed tunable luminaire \cite{alrashdi2023}, which is used to generate known light spectra, as depicted in \ref{fig5}(b).

The rest of this paper is organized as follows: Section II details the methodology used to model the individual components of the camera pipeline, along with a review of relevant literature. Section III presents the experimental results and observations. Section IV concludes the paper.

\section{Methodology}

\subsection{Pixel Intensity Estimation}

The physics of image acquisition is depicted in \ref{fig3}. Here, $l(\lambda)$ represents the spectral power distribution of the illuminant, and $r(\lambda)$ denotes the spectral reflectance of the object surface as functions of wavelength $\lambda$. The scene radiance observed by the camera is $p(\lambda)\in\mathbb{R}^{1\times{M}}$ defined for $M$ discrete wavelengths, resulting from the interaction of $l(\lambda)$ and $r(\lambda)$. The camera’s response is a pixel intensity triplet, each corresponding to the red, green, and blue channels. Generally the intensities are represented as a vector $I=[I_r,I_g,I_b]$.

The pixel intensity can be estimated by:
\begin{equation}
I = g(h(rl\Omega))\label{eq1}
\end{equation}

Throughout this manuscript, we omit explicit references to wavelength ($\lambda$) for brevity. Additionally, channel indices (e.g. $k$ in $I_{k}$) are omitted where convenient. The terms $\Omega\in\mathbb{R}^{M\times{3}}$, $g$, and $h$ represent the camera’s spectral sensitivity function, response function, and gamut mapping function, respectively. The response function $g$ consists of three channel-specific functions, $g_r$, $g_g$, and $g_b$, which correspond to the red, green, and blue channels, respectively.

Each of these functions plays a critical role in determining how the camera processes scene radiance. The next subsections explore these functions in detail, highlighting their importance in constructing an accurate camera model.

\subsection{Pixel Intensity Saturation}

\begin{figure}[h]
\includegraphics[width=\columnwidth]{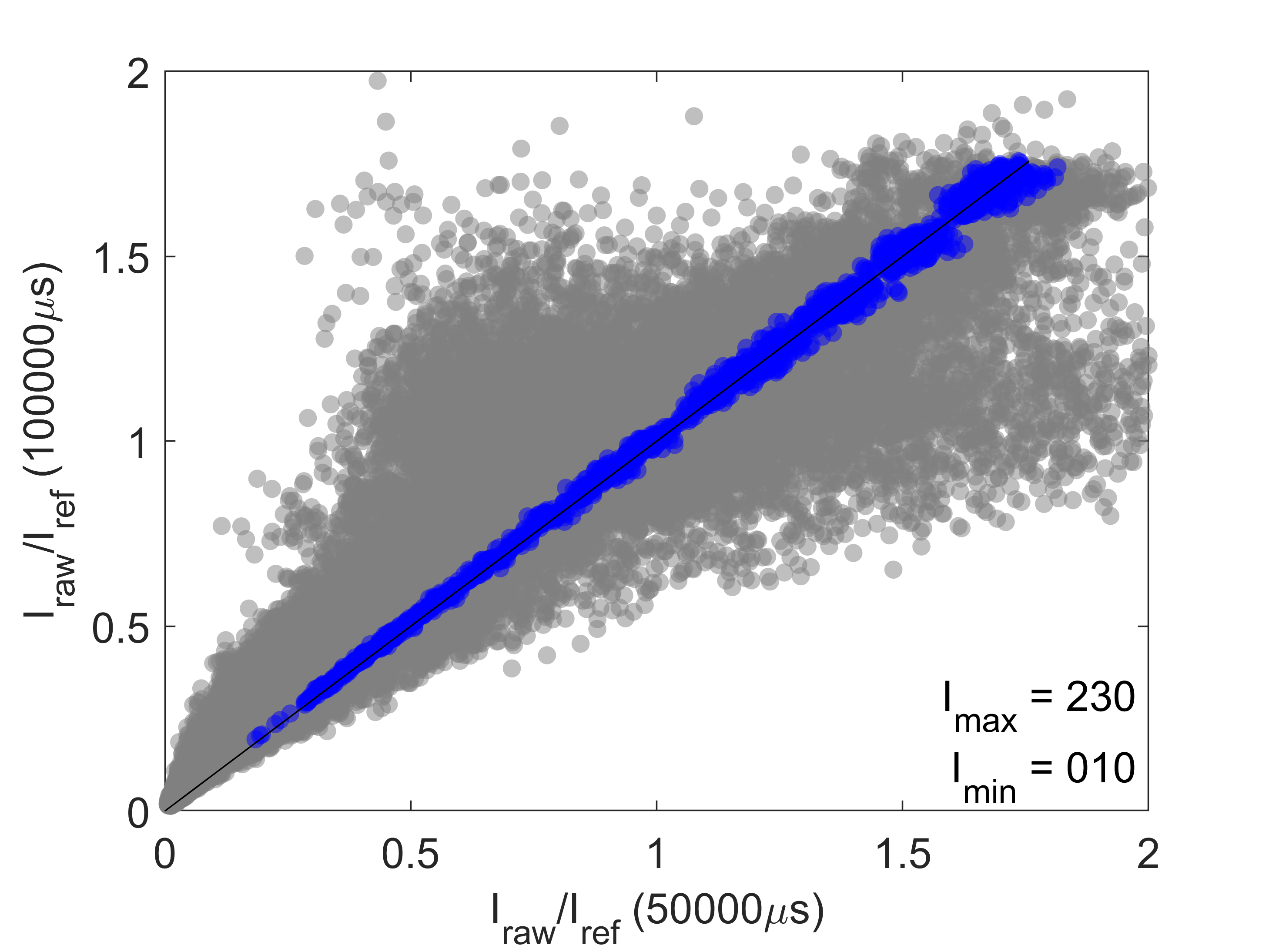}
\caption{Plot of pixel intensity in the blue camera channel at different exposure times. Gray-shaded pixels indicate saturation in at least one of the other camera channels. Pixel intensities below 10 or above 230 are classified as saturated.}
\label{fig4}
\end{figure}

The effects of pixel intensity saturation have been examined by Kim \textit{et al.} \cite{kim2012}, who demonstrated that the saturation of a single camera channel influences all other channels. As illustrated in Fig.~\ref{fig4}, this interaction adds complexity to the behavior of saturated pixels, which follow a non-linear mapping that is challenging to estimate accurately.

\subsection{Camera Response}

The camera response is modeled as the final element in the imaging workflow. Early research, such as by Chang \textit{et al.} \cite{chang1996}, treated the nonlinearity of the response as a pixel intensity error which was effectively modeled and corrected out of the image. Later work by Vora and colleagues \cite{vora1997a, vora1997b} conceptualized the response as a standalone function. However, their approach lacks sufficient definition for systematic application across diverse imaging systems.

The response function was comprehensively defined and analyzed by Nayar and colleagues \cite{mitsunaga1999, grossberg2003, grossberg2004}. They demonstrated that the ratio of pixel intensities at different exposure settings corresponds to the ratio of intensities prior to the application of the camera response function:

\begin{equation}
\frac{I_{k,e_1}}{I_{k,e_2}} = \frac{g(E_{k,e_1})}{g(E_{k,e_2})} = \frac{e_1}{e_2}\label{eq2}
\end{equation}

Here, $I_{k,e_1}$ represents the pixel intensity of the $k$\textsuperscript{th} camera channel ($k$ being red, green, or blue) at an exposure time $e_1$. Given measurements of $I_{k,e_1}$ and $I_{k,e_2}$ at known $e_1$ and $e_2$, and assuming $g$ is a polynomial transformation between $I$ and $E$, the response function $g$ can be derived. In later work, Grossberg and Nayar \cite{grossberg2004} provided a database of known camera response functions, which can be utilized to estimate the response of new cameras. More recent work \cite{toivonen2020} uses a specialized setup to generate the response function.

In our experiments, we have found that the more simple formulation introduced by Debevec and Malik \cite{debevec1997} give the best results.

\begin{align}
I_{k,e_i} &= g(E_{k}e_i)\nonumber\\
\ln{\left( g^{-1}(I_{k,e_i}) \right)} &= \ln{E_{k}} + \ln{e_i}\label{eq3}
\end{align}

Here, only one set of measurements of $I_{k,e_i}$ are required at a known exposure $e_i$.

\subsection{Camera Sensitivity}

The spectral sensitivity of a camera defines the efficiency of the imaging sensor in converting input scene radiance at specific wavelengths into electrical signals. To extend the operational range within the visible spectrum, modern cameras are typically equipped with infrared (IR) filters. Consequently, the camera’s effective spectral sensitivity is a combination of the sensor’s intrinsic sensitivity and the spectral filtering characteristics of the IR filter.

\begin{figure*}[t]
\includegraphics[width=\textwidth]{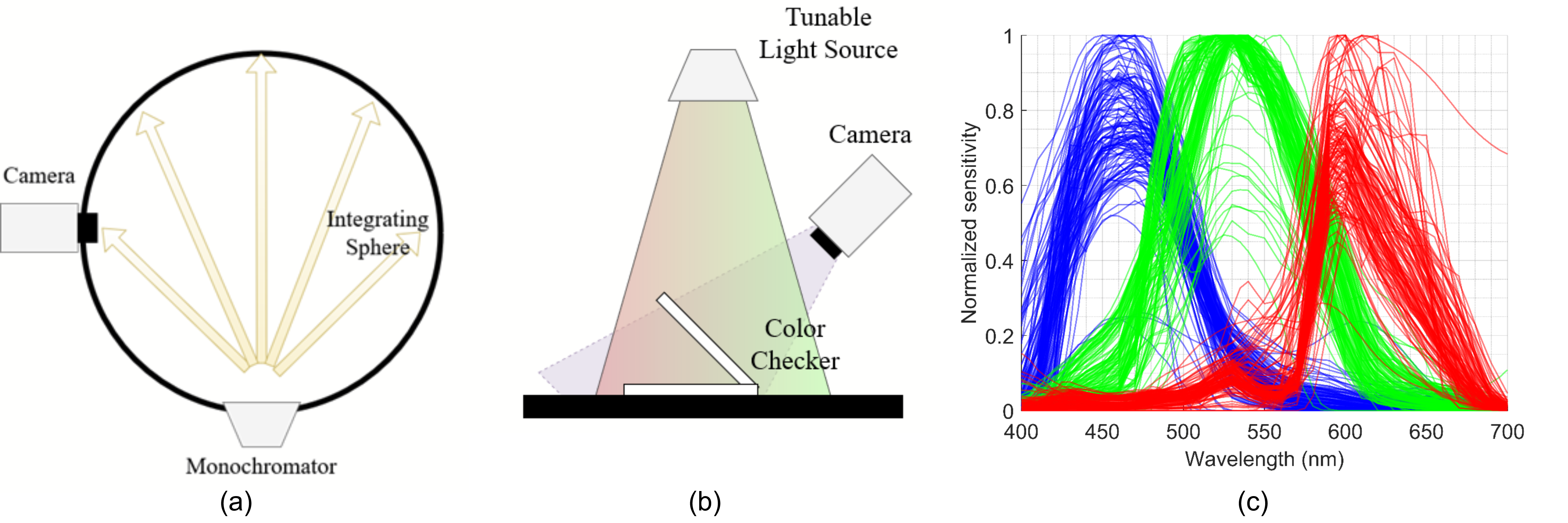}
\caption{Camera Sensitivity Estimation. (a) The traditional method of sensitivity estimation, involving an integrating sphere and a monochromator. (b) The estimation technique employed in this investigation, which uses images of a known target under controlled illumination. (c) A database of known camera sensitivity functions, used for basis decomposition.}
\label{fig5}
\end{figure*}

Traditionally, sensitivity functions are accurately estimated using specialized optical equipment, as illustrated in Fig.~\ref{fig5}(a). In this setup, a monochromator and a camera are attached to an integrating sphere, allowing the camera's sensitivity at each wavelength to be determined by varying the light output from the monochromator \cite{berra2015}. However, this technique is both restrictive and expensive, rendering it inaccessible to many research groups. An excellent overview of other early techniques is provided in \cite{hubel1994}.

Fig.~\ref{fig5}(b) depicts a method developed in later investigations. In this approach, images of a target with known reflectance, such as a color checker, are captured under a known light source. The resulting images are then used to estimate the camera spectral sensitivity function. Early work by Hardeberg \textit{et al.} \cite{hardeberg1998} employed a direct pseudo-inverse of the linear pixel intensity formula to derive $\Omega_k$:

\begin{equation}
\Omega_{k} = (P^{T}P)^{-1}P^{T}I_{k}\label{eq4}
\end{equation}

Here, $\Omega_k$ represents the spectral sensitivity function for the $k$\textsuperscript{th} camera channel, and $P_{k}\in\mathbb{R}^{N\times{M}}$ represents a set of $N$ spectral measurements. However, direct pseudo-inversion is unreliable as it is highly susceptible to noise. To address this limitation, later studies \cite{zhao2009, kang2011} introduced regularization by incorporating basis vectors. Subsequent modifications to this approach were proposed in \cite{han2012, rei2013, zhu2020}. Finlayson \textit{et al.} \cite{finlayson2016} presented a rank-based estimation method, which identifies the possible space of the sensitivity function based on an initial estimate by a process of deduction by elimination.

We found the best technique to be the one presented by Jiang \textit{et al.} \cite{jiang2013}, which demonstrates that the spectral sensitivity estimates of different cameras are inherently low-dimensional. This insight enables the estimation of new camera sensitivities using a constrained regularization approach based on known spectral sensitivity functions.

As shown in Fig.~\ref{fig5}(c), we assembled a database of spectral sensitivity functions for 126 cameras sourced from various online datasets and publications \cite{jiang2013,zhao2009,tominaga2021,darrodi2015,iedata}. This database was decomposed into three basis matrices, $B_k$, where each $k$ corresponds to one camera channel. The spectral sensitivity of an unknown camera was then estimated using a constrained least-squares solution, as described in (\ref{eq5}):

\begin{align}
A_k &= P_{k}B_{k}\nonumber\\
c_k &= (A_k^{T}A_k)^{-1}A_k^{T}P_k \quad \textrm{s.t.} \quad c_kB_k\geq0\nonumber\\
\hat{\Omega}_k &= \operatorname{diag}(c_{k}B_{k})\label{eq5}
\end{align}

\begin{figure}[h]
\includegraphics[width=\columnwidth]{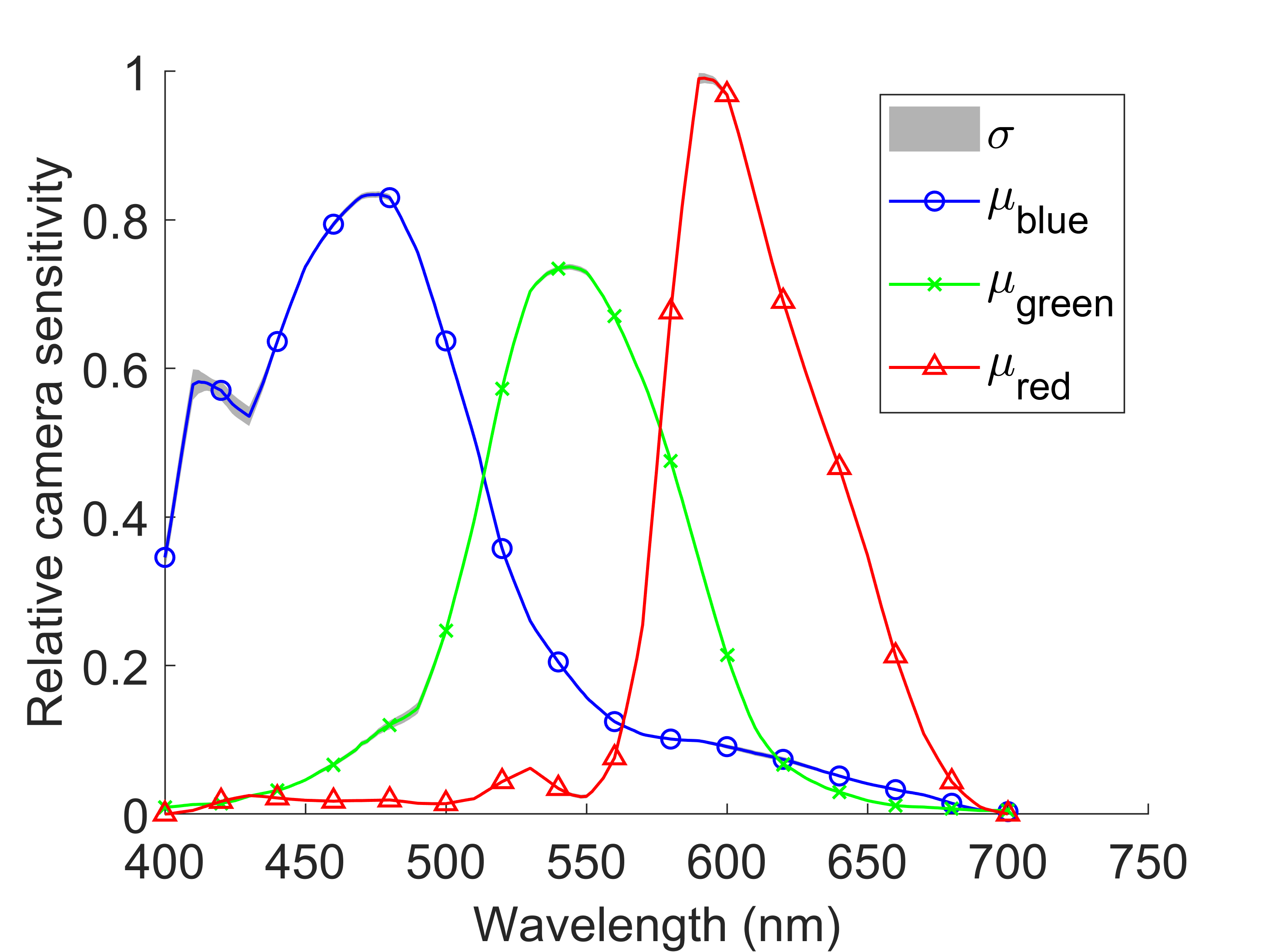}
\caption{Estimated camera spectral sensitivity function. Here, $\sigma$ represents the standard deviation from a 10-fold cross-validation, while ${\mu}_k$ denotes the mean sensitivity function.}
\label{fig6}
\end{figure}

The hat notation, $\hat{\Omega}_k$, indicates that this is an estimate of the spectral sensitivity function, $\Omega_k$. The constraints ensure that $\hat{\Omega}_k$ remains non-negative. For the sensitivity estimation, only unsaturated pixels, as illustrated in Fig.~\ref{fig4} are used. The resulting sensitivity for the camera used in this investigation is illustrated in Fig.~\ref{fig6}.

\subsection{Gamut Mapping}

\begin{figure}[h]
\includegraphics[width=\columnwidth]{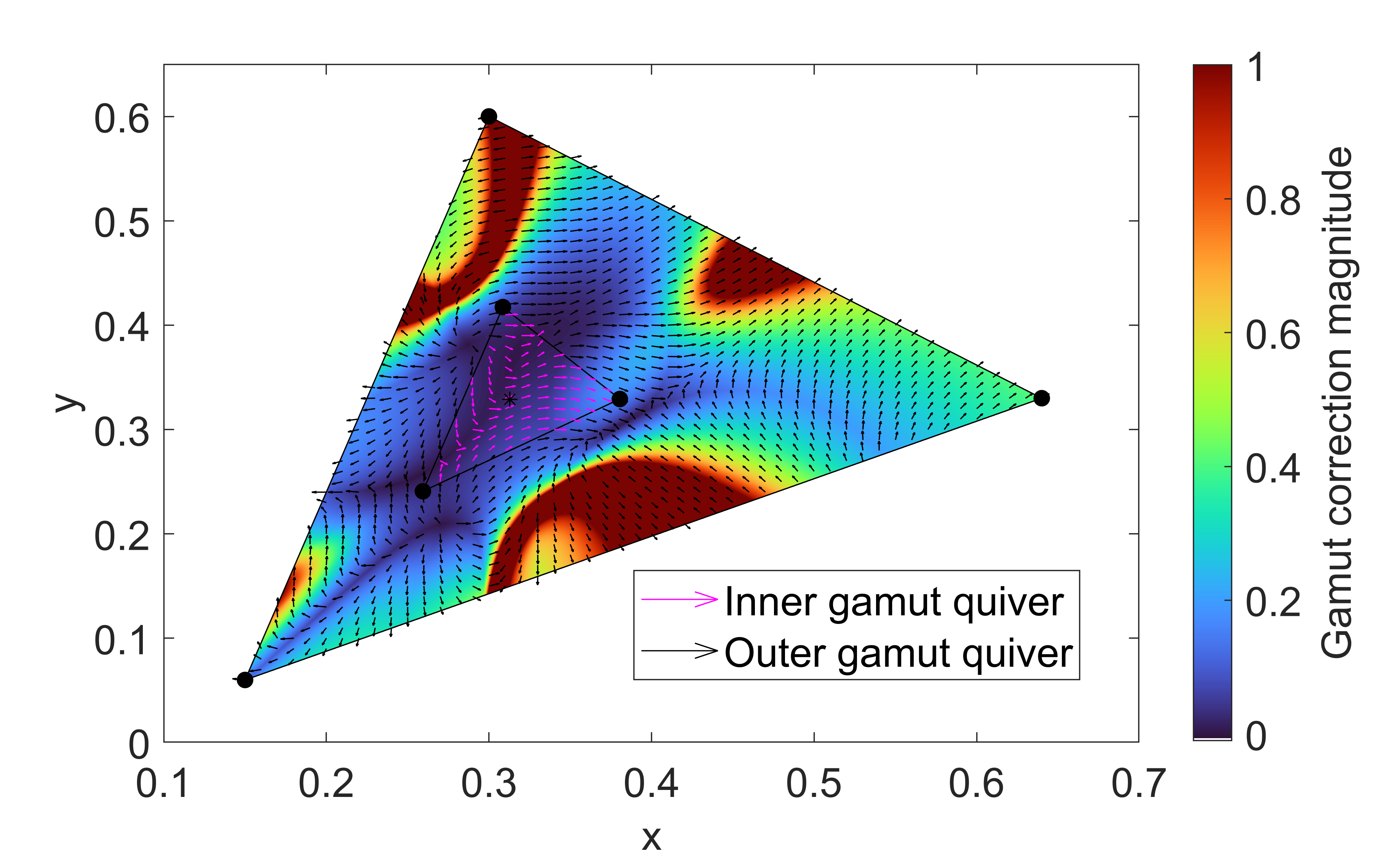}
\caption{The gamut mapping procedure, with pixel values projected onto the sRGB color gamut in the CIE 1931 chromaticity diagram. The quivers show the direction of the transformation from $S$ to $E$, while the heatmap indicates the magnitude of the change.}
\label{fig7}
\end{figure}

The gamut mapping function has been extensively studied by Kim \textit{et al.} \cite{kim2012}. It describes the transformation between $S$ and $E$ in the sRGB color space. This procedure is illustrated in Fig.~\ref{fig7}. Since both $\Omega$ and $g$ have already been estimated, $S$ and $E$ are known, allowing the transformation function $h$ to be computed as $E = h(S)$.

As shown in Fig.~\ref{fig7}, both $S$ and $E$ are mapped onto the sRGB color gamut in the CIE 1931 chromaticity diagram. The quivers represent the transformation direction (from $S$ to $E$), while the heatmap illustrates the magnitude of the change.

The gamut mapping function varies significantly between cameras, leading to more vivid images in some cases and darker images in others. Due to its highly nonlinear nature, developing a universal parametric model is challenging. As a solution, Kim \textit{et al.} \cite{kim2012} propose using Radial Basis Functions (RBFs) for approximation, which we have found to be the most effective approach.

\section{Results and Discussion}

\begin{figure}[h]
\includegraphics[width=\columnwidth]{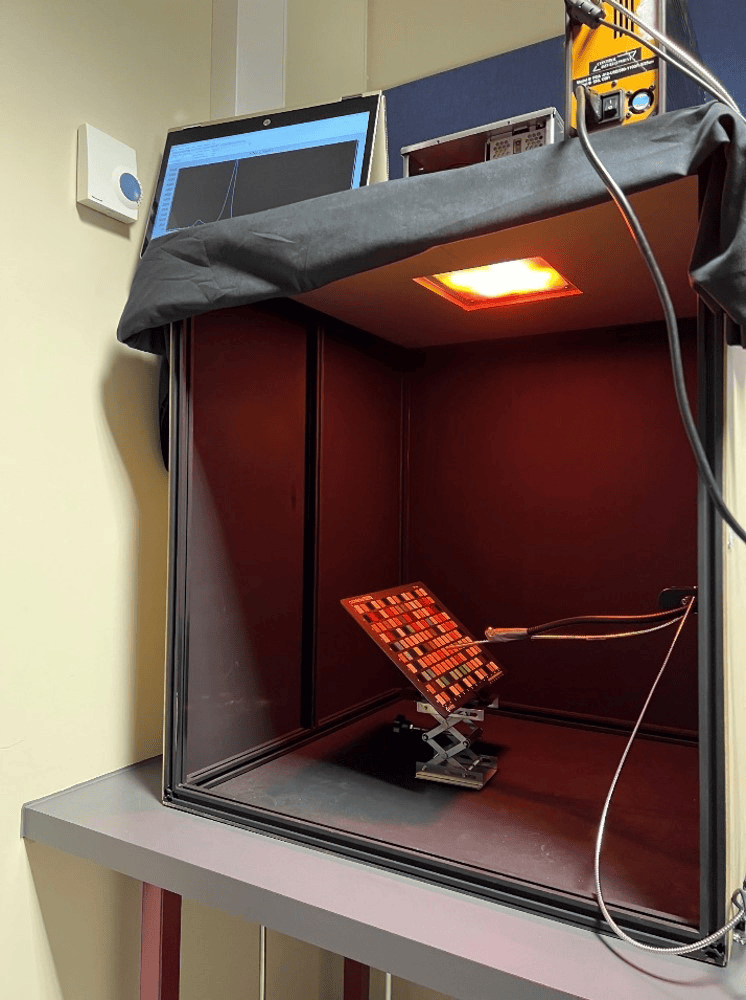}
\caption{The experimental setup.}
\label{fig8}
\end{figure}

This investigation employs a Logitech C920 HD Pro 8-bit web camera to demonstrate the proposed pipeline. The experimental setup, shown in Fig.~\ref{fig8}, includes a custom-designed programmable light source used to generate multiple illumination spectra \cite{alrashdi2023}. Validation data for testing the camera model was obtained by imaging a ColorChecker SG under various illumination conditions.

\begin{figure}[h]
\includegraphics[width=\columnwidth]{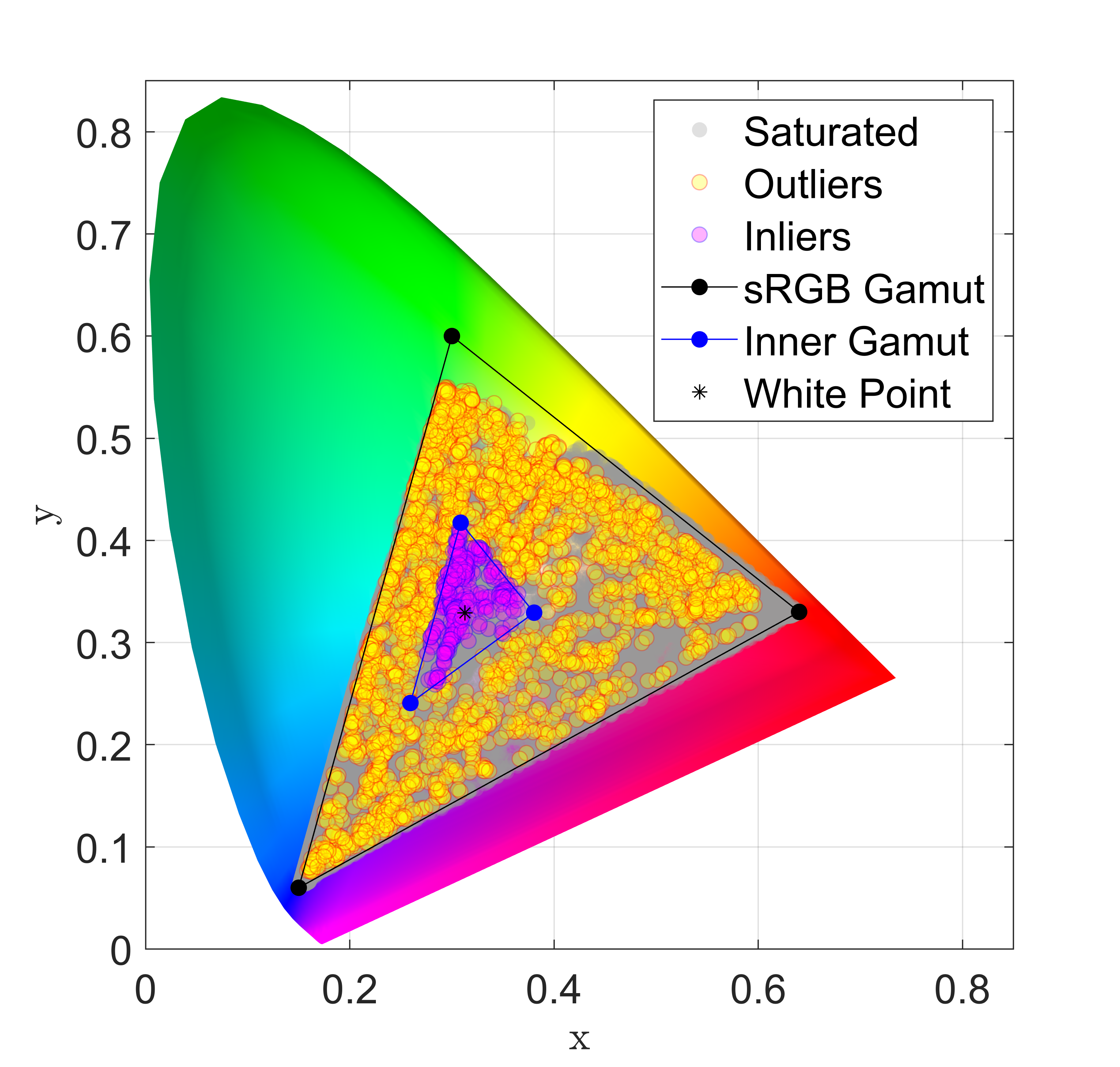}
\caption{Measured pixel intensity data projected onto the CIE 1931 chromaticity diagram. The data is divided into an inner and outer gamut, with the inner gamut data exhibiting minimal changes due to gamut mapping.}
\label{fig9}
\end{figure}

It is challenging to simultaneously estimate all three functions $h$, $g$, and $\Omega$. Therefore, the estimation is carried out in two stages. The first stage ignores the effects of the gamut mapping function, as Kim \textit{et al.} \cite{kim2012} demonstrated that its impact is minimal at the central region of the sRGB gamut. As illustrated in Fig.~\ref{fig9}, the captured pixel intensity data are projected onto the CIE 1931 chromaticity diagram and divided into two regions: the inner gamut and outer gamut. The size of the inner gamut is not strictly defined; in this study, we define it to ensure a sufficient number of data points remain on the inner gamut region for accurate estimation.

\begin{figure}[h]
\includegraphics[width=\columnwidth]{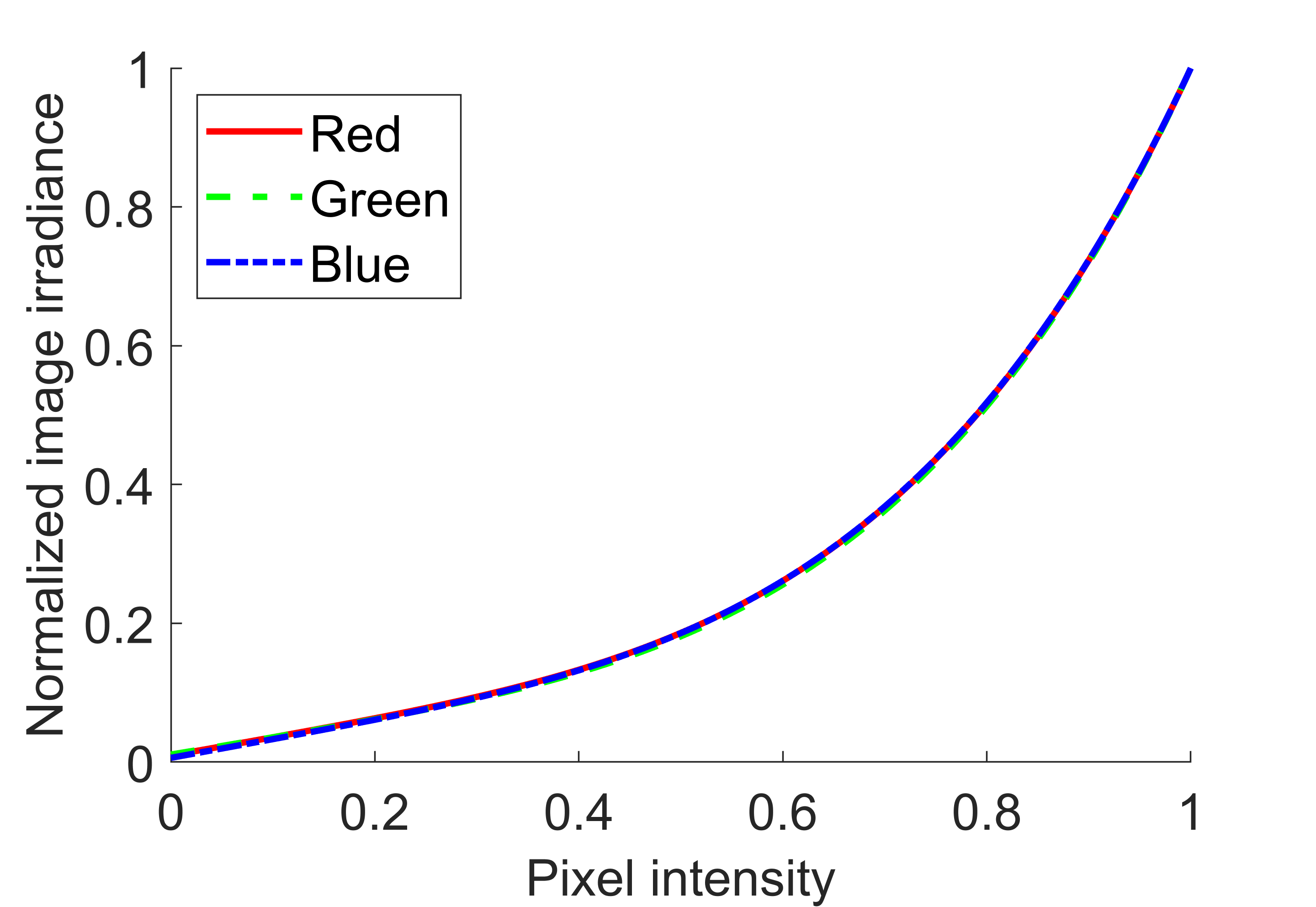}
\caption{Estimated camera response function, derived from inner gamut data.}
\label{fig10}
\end{figure}

In the first stage, only inner gamut data is used to estimate the camera sensitivity function and response function, with the resulting response function shown in Fig.~\ref{fig10}. In the second stage, data from both the inner and outer gamut is used to estimate the gamut mapping function.

\begin{figure}[h]
\includegraphics[width=\columnwidth]{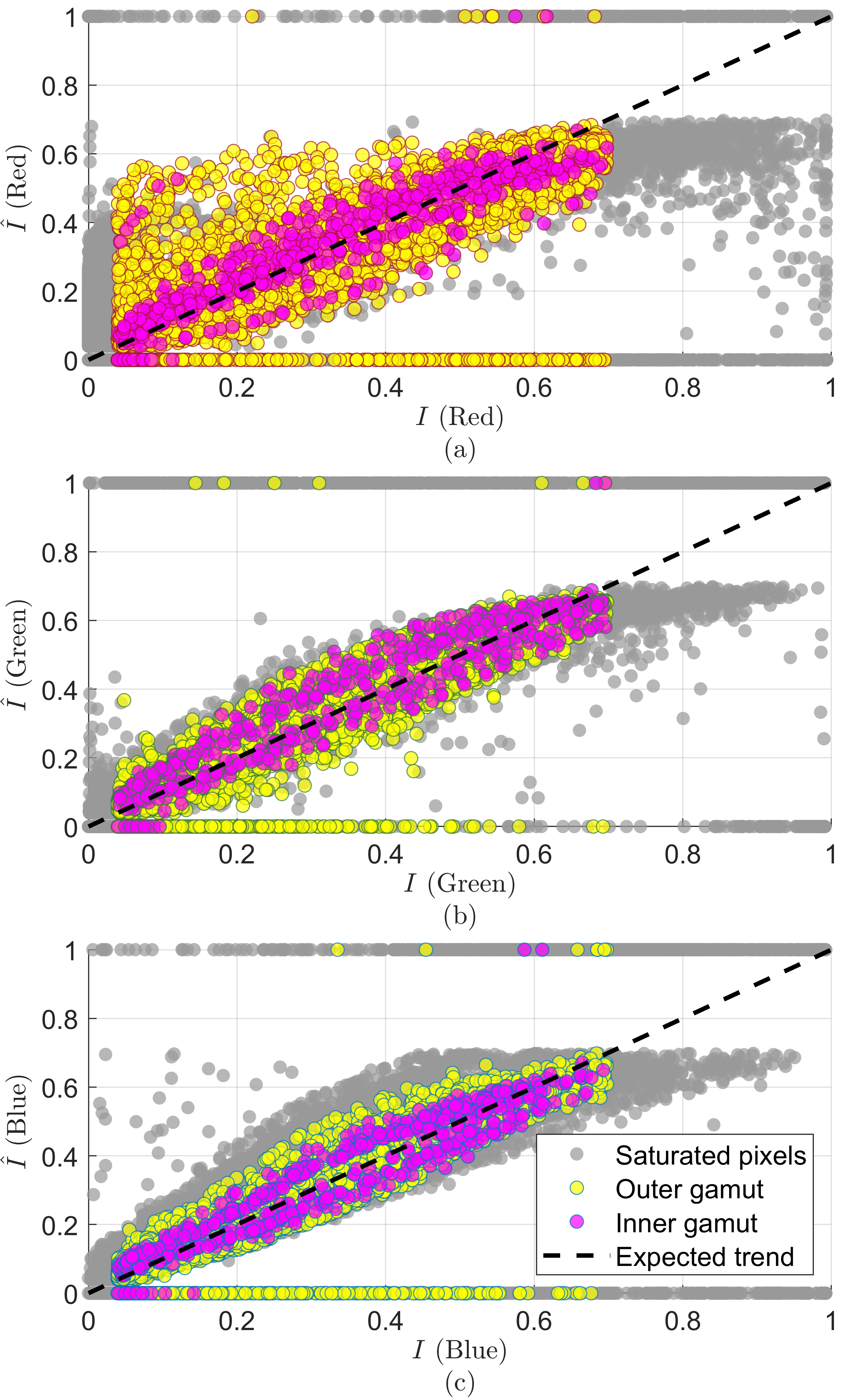}
\caption{The estimated ($\hat{I}$) vs measured ($I$) pixel intensity for the (a) red, (b) green, and (c) blue camera channels.}
\label{fig11}
\end{figure}

The estimation accuracy for the three camera channels are illustrated in Fig.~\ref{fig11}. The results indicate a linear response for unsaturated pixels. However, saturated pixels exhibit a larger spread, even after modeling. This suggests that an additional modeling step may be necessary or that the gamut mapping procedure requires further refinement.

\section{Conclusion}

This study presents a comprehensive pipeline for estimating the pixel intensity of a camera. By leveraging a database of known spectral sensitivity functions, response functions, and gamut mapping techniques, the proposed approach addresses key challenges in camera characterization, including noise susceptibility and nonlinear transformations. Validation with a custom programmable light source demonstrated accurate modeling for unsaturated pixels, though challenges remain with saturated pixels, which exhibit greater variability. These findings highlight the need for further refinement in gamut mapping and additional modeling steps to enhance accuracy. The proposed approach offers a practical framework for camera characterization, with significant potential applications in computational photography and color science.

\section*{Acknowledgment}

This work was supported by the Royal Society of New Zealand through a Rutherford Discovery Fellowship awarded to Ooi, M. P.-L., for the project Resilient and Efficient Light-Based Plant Detection and Characterization for Precision Agriculture and Environmental Sustainability (grant RDF-UOW1902).

\vspace{12pt}

\bibliographystyle{IEEEtran}
\bibliography{bibliography}

% Generated by IEEEtran.bst, version: 1.14 (2015/08/26)
\begin{thebibliography}{10}
\providecommand{\url}[1]{#1}
\csname url@samestyle\endcsname
\providecommand{\newblock}{\relax}
\providecommand{\bibinfo}[2]{#2}
\providecommand{\BIBentrySTDinterwordspacing}{\spaceskip=0pt\relax}
\providecommand{\BIBentryALTinterwordstretchfactor}{4}
\providecommand{\BIBentryALTinterwordspacing}{\spaceskip=\fontdimen2\font plus
\BIBentryALTinterwordstretchfactor\fontdimen3\font minus \fontdimen4\font\relax}
\providecommand{\BIBforeignlanguage}[2]{{%
\expandafter\ifx\csname l@#1\endcsname\relax
\typeout{** WARNING: IEEEtran.bst: No hyphenation pattern has been}%
\typeout{** loaded for the language `#1'. Using the pattern for}%
\typeout{** the default language instead.}%
\else
\language=\csname l@#1\endcsname
\fi
#2}}
\providecommand{\BIBdecl}{\relax}
\BIBdecl

\bibitem{rodrigues2020}
P.~Rodrigues, J.~Barreto, and M.~Antunes, ``Photometric camera characterization from a single image with invariance to light intensity and vignetting,'' \emph{Computer Vision and Image Understanding}, vol. 192, p. 102887, 2020.

\bibitem{grossberg2004}
M.~Grossberg and S.~Nayar, ``Modeling the space of camera response functions,'' \emph{IEEE Transactions on Pattern Analysis and Machine Intelligence}, vol.~26, no.~10, pp. 1272--1282, 2004.

\bibitem{asada1996}
N.~Asada, A.~Amano, and M.~Baba, ``Photometric calibration of zoom lens systems,'' in \emph{Proceedings of 13th International Conference on Pattern Recognition}, vol.~1, 1996, pp. 186--190.

\bibitem{klein1996}
S.~Klein, Q.~Hu, and T.~Carney, ``The adjacent pixel nonlinearity: Problems and solutions,'' \emph{Vision Research}, vol.~36, no.~19, pp. 3167--3181, 1996.

\bibitem{debevec1997}
P.~Debevec and J.~Malik, ``Recovering high dynamic range radiance maps from photographs,'' in \emph{Proceedings of the 24th Annual Conference on Computer Graphics and Interactive Techniques}, ser. SIGGRAPH '97.\hskip 1em plus 0.5em minus 0.4em\relax USA: ACM Press/Addison-Wesley Publishing Co., 1997, p. 369–378.

\bibitem{kim2012}
S.~Kim, H.~Lin, Z.~Lu, S.~Süsstrunk, S.~Lin, and M.~Brown, ``A new in-camera imaging model for color computer vision and its application,'' \emph{IEEE Transactions on Pattern Analysis and Machine Intelligence}, vol.~34, no.~12, pp. 2289--2302, 2012.

\bibitem{alrashdi2023}
A.-R. Mohammed, S.~Abeysekera, M.-L. Ooi, Y.-C. Kuang, V.~Kalavally, and C.~Cheng, ``Light driven visual inspection system for human vision,'' in \emph{2023 IEEE International Instrumentation and Measurement Technology Conference (I2MTC)}, 2023, pp. 1--6.

\bibitem{chang1996}
Y.-C. Chang and J.~Reid, ``Rgb calibration for color image analysis in machine vision,'' \emph{IEEE Transactions on Image Processing}, vol.~5, no.~10, pp. 1414--1422, 1996.

\bibitem{vora1997a}
P.~Vora and J.~Farrell, ``Digital color cameras - 2 - spectral response,'' 1997.

\bibitem{vora1997b}
P.~Vora, J.~Farrel, T.~Jerome, and D.~Brainard, ``Linear models for digital cameras,'' 04 1997.

\bibitem{mitsunaga1999}
T.~Mitsunaga and S.~Nayar, ``Radiometric self calibration,'' in \emph{Proceedings of 1999 IEEE Computer Society Conference on Computer Vision and Pattern Recognition}, vol.~1, 1999, pp. 374--380 Vol. 1.

\bibitem{grossberg2003}
M.~Grossberg and S.~Nayar, ``Determining the camera response from images: what is knowable?'' \emph{IEEE Transactions on Pattern Analysis and Machine Intelligence}, vol.~25, no.~11, pp. 1455--1467, 2003.

\bibitem{toivonen2020}
M.~Toivonen and A.~Klami, ``Practical camera sensor spectral response and uncertainty estimation,'' \emph{Journal of Imaging}, vol.~6, no.~8, 2020.

\bibitem{berra2015}
E.~Berra, S.~Gibson-Poole, A.~Macarthur, R.~Gaulton, and A.~Hamilton, ``Estimation of the spectral sensitivity functions of un-modified and modified commercial off-the-shelf digital cameras to enable their use as a multispectral imaging system for uavs,'' in \emph{ISPRS - International Archives of the Photogrammetry, Remote Sensing and Spatial Information Sciences}, vol. XL-1/W4, 08 2015.

\bibitem{hubel1994}
P.~Hubel, D.~Sherman, and J.~Farrell, ``A comparison of methods of sensor spectral sensitivity estimation,'' in \emph{International Conference on Communications in Computing}, 1994.

\bibitem{hardeberg1998}
J.~Hardeberg, H.~Brettel, and F.~Schmitt, ``Spectral characterization of electronic cameras,'' in \emph{Electronic Imaging: Processing, Printing, and Publishing in Color}, J.~Bares, Ed., vol. 3409, International Society for Optics and Photonics.\hskip 1em plus 0.5em minus 0.4em\relax SPIE, 1998, pp. 100--109.

\bibitem{zhao2009}
H.~Zhao, K.~Rei, R.~Tan, and K.~Ikeuchi, ``Estimating basis functions for spectral sensitivity of digital cameras,'' 01 2009.

\bibitem{kang2011}
M.~Kang, U.~Yang, and K.~Sohn, ``Spectral sensitivity estimation for emccd camera,'' \emph{Electronics Letters}, vol.~47, pp. 1369--1370, 12 2011.

\bibitem{han2012}
S.~Han, Y.~Matsushita, I.~Sato, T.~Okabe, and Y.~Sato, ``Camera spectral sensitivity estimation from a single image under unknown illumination by using fluorescence,'' in \emph{2012 IEEE Conference on Computer Vision and Pattern Recognition}, 2012, pp. 805--812.

\bibitem{rei2013}
K.~Rei, H.~Zhao, R.~Tan, and K.~Ikeuchi, ``Camera spectral sensitivity and white balance estimation from sky images,'' \emph{International Journal of Computer Vision}, vol. 105, 12 2013.

\bibitem{zhu2020}
J.~Zhu, X.~Xie, N.~Liao, Z.~Zhang, W.~Wu, and L.~Lv, ``Spectral sensitivity estimation of trichromatic camera based on orthogonal test and window filtering,'' \emph{Opt. Express}, vol.~28, no.~19, pp. 28\,085--28\,100, 09 2020.

\bibitem{finlayson2016}
G.~Finlayson, M.~Darrodi, and M.~Mackiewicz, ``Rank-based camera spectral sensitivity estimation,'' \emph{J. Opt. Soc. Am. A}, vol.~33, no.~4, pp. 589--599, 04 2016.

\bibitem{jiang2013}
J.~Jiang, D.~Liu, J.~Gu, and S.~Süsstrunk, ``What is the space of spectral sensitivity functions for digital color cameras?'' in \emph{2013 IEEE Workshop on Applications of Computer Vision (WACV)}, 2013, pp. 168--179.

\bibitem{tominaga2021}
S.~Tominaga, S.~Nishi, and R.~Ohtera, ``Measurement and estimation of spectral sensitivity functions for mobile phone cameras,'' \emph{Sensors}, vol.~21, no.~15, 2021.

\bibitem{darrodi2015}
M.~Darrodi, G.~Finlayson, T.~Goodman, and M.~Mackiewicz, ``Reference data set for camera spectral sensitivity estimation,'' \emph{J. Opt. Soc. Am. A}, vol.~32, no.~3, pp. 381--391, 03 2015.

\bibitem{iedata}
I.~Engineering, ``Data and tools library,'' \url{https://www.image-engineering.de/library/data-and-tools}, 2024.

\end{thebibliography}

\end{document}